# Multilingual Schema Matching for Wikipedia Infoboxes

Thanh Nguyen[1]   Viviane Moreira[2]   Huong Nguyen[1]   Hoa Nguyen[1]   Juliana Freire[3]

[1]University of Utah    [2]UFRGS-Brazil    [3]NYU Poly

{thanhnh,huongnd,thanhhoa}@cs.utah.edu    viviane@inf.ufrgs.br    juliana.freire@nyu.edu

## ABSTRACT

Recent research has taken advantage of Wikipedia's multilingualism as a resource for cross-language information retrieval and machine translation, as well as proposed techniques for enriching its cross-language structure. The availability of documents in multiple languages also opens up new opportunities for querying structured Wikipedia content, and in particular, to enable answers that straddle different languages. As a step towards supporting such queries, in this paper, we propose a method for identifying mappings between attributes from infoboxes that come from pages in different languages. Our approach finds mappings in a completely automated fashion. Because it does not require training data, it is scalable: not only can it be used to find mappings between many language pairs, but it is also effective for languages that are under-represented and lack sufficient training samples. Another important benefit of our approach is that it does not depend on syntactic similarity between attribute names, and thus, it can be applied to language pairs that have distinct morphologies. We have performed an extensive experimental evaluation using a corpus consisting of pages in Portuguese, Vietnamese, and English. The results show that not only does our approach obtain high precision and recall, but it also outperforms state-of-the-art techniques. We also present a case study which demonstrates that the multilingual mappings we derive lead to substantial improvements in answer quality and coverage for structured queries over Wikipedia content.

## 1. INTRODUCTION

With over 17.9 million articles and 10 million page views per month [38], Wikipedia has become a popular and important source of information. One of its most remarkable aspects is *multilingualism*: there are Wikipedia articles in over 270 languages. This opens up new opportunities for knowledge sharing among people that speak different languages both within and outside the scope Wikipedia. For example, cross-language links, that connect an article in one language to the corresponding article in another, have been used to derive better translations in cross-language information retrieval and machine translation [11, 24, 30, 32].

But even though many languages are represented in Wikipedia, the geographical distribution of Wikipedia users is highly skewed. One of the explanations for this effect is that many languages, including languages spoken by large segments of the world population, are under-represented. For example, there are 328 million English speakers worldwide and 20% of the Wikipedia pages are in English; in contrast, there are 178 million Portuguese speakers and only 3.75% of Wikipedia articles are in Portuguese. Recognizing this problem, there are a number of ongoing efforts which aim to improve access to Wikipedia content. By leveraging the existing multilingual Wikipedia corpus, techniques have been proposed to: combine content provided in documents from different languages and thereby improve both documents [1, 5]; find missing cross-language links [29, 33]; aid in the creation of multilingual content [19]; and help users who speak different languages to search for named entities in the English Wikipedia [35].

Besides textual content, Wikipedia has also become a prominent source for structured information. A growing number of articles contain an *infobox* that provides a structured record for the entity described in the article. This has enabled richer queries over Wikipedia content (see *e.g.,* [2, 17, 25]). While much work has been devoted to supporting structured queries, no previous effort has looked into providing support for multilingual structured queries. In this paper, we examine the problem of *matching schemas of infoboxes represented in different languages*, a necessary step for supporting these queries.

By discovering multilingual attribute correspondences, it is possible to integrate information from different languages and to provide more complete answers to user queries. A common scenario is when the answer to a query cannot be found in a given language but it is available in another. In a study of the 50 topics used in the GikiCLEF campaign [13], just nine topics had answers in all ten languages used in the task [6]. However, almost every query had an answer in the English Wikipedia. Thus, by supporting multi-language queries and providing the relevant English documents as part of the answer, recall can be improved for most other languages. In addition, some queries can benefit from integrating information present in multiple infoboxes represented in different languages. Consider the query *Find the genre and the studio that produced the film "The Last Emperor"*. To provide a complete answer to this query, we need to combine the information from the two infoboxes in Figure 1.

There are several challenges involved in finding multilingual correspondences across infoboxes. Even within a language, finding attribute correspondences is difficult. Although authors are encouraged to provide structure in Wikipedia articles, *e.g.,* by selecting appropriate templates and categories, they often do not follow the guidelines or follow them loosely. This leads to several problems, in particular, schema drift—the structure of infoboxes for the same entity type (*e.g.,* actor, country) can differ for different instances. Both polysemy and synonymy are observed among attribute names: a given name can have different semantics (*e.g., born* can mean *birth date* or *country of birth*) and different names can have the same meaning (*e.g., alias* and *other names*). This problem is compounded when we consider multiple languages. Figure 1 shows an example of heterogeneity in infoboxes describing the same entity in different languages. Some attributes in the English infobox do not have a counterpart in the Portuguese infobox and vice-versa. For instance: *produced by, editing by, distributed by,* and *budget* are omitted in the Portuguese version, while *género* (genre) is omitted in the English version. An analysis of the overlap among attribute sets from infoboxes in English and Portuguese (see Table 5) shows that on average only 42% of the attributes are present in both languages. Besides the variation in structure, there are also inconsistencies in the attribute values, for example: *running time* is 160 minutes in the English version and 165 minutes in the Portuguese version; Ryuichi Sakamoto appears under *Music by* in English and under *Elenco original* (cast) in Portuguese.

To identify multilingual matches, a possible strategy would be to translate the attribute names and values using a multilingual dictionary or a machine translation system, and then apply traditional schema or ontology matching techniques [31, 10, 12]. However, this strategy is limited since, in many cases, the correct correspondence is not found among the translations. For example, in articles describing movies, the correct alignment for the English attribute *starring* is the Portuguese attribute *elenco original*. However, the dictionary translation is *estrelando* for the former and *original cast* for the latter, and neither is used in the Wikipedia infobox templates to name an attribute. WordNet is another source of synonyms that can potentially help in matching, but its versions in many languages are incomplete. For instance, the Vietnamese WordNet [36] covers only 10% of the senses present in the English WordNet. Furthermore, traditional techniques such as string similarity may fail even for languages that share words with similar roots. Consider the term *editora*, which in Portuguese means *publisher*. Using string similarity, it would be very close to *editor*, but this would be a false cognate.

Recently, techniques have been proposed to identify multilingual attribute alignments for Wikipedia infoboxes. But these have important shortcomings in that they are designed for languages that share similar words [1, 5], or demand a considerable amount of training data [1]. Consequently, they cannot be effectively applied to languages with distinct representations or different roots; and their applicability is also limited for under-represented languages in Wikipedia, which have few pages and thus, insufficient training data.

**Contributions.** We propose `WikiMatch`, a new approach to multilingual schema matching that addresses these limitations. `WikiMatch` gathers similarity evidence from multiple sources: attribute values, link structure, co-occurrence statistics within and across languages, and an automatically derived bilingual dictionary. These different sources of similarity information are combined in a systematic manner: the alignment algorithm prioritizes the derivation of high-confidence correspondences and then uses these to find additional ones. By doing so, it is able to obtain both high precision and recall. The algorithm finds, in a single step, inter- and intra-language correspondences, as well as complex, one-to-many correspondences. Because `WikiMatch` does not require training data, it is able to handle under-represented languages; and since it does not rely on string similarity on attribute names, it can be applied both to similar and morphologically distinct languages. Furthermore, it does not require external resources, such as bilingual dictionaries, thesauri, ontologies, or automatic translators.

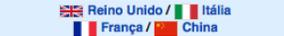

(a) English  (b) Portuguese

**Figure 1: Excerpts from English and Portuguese infoboxes for the Film *The Last Emperor*.**

We present a detailed experimental evaluation using infoboxes in Portuguese, Vietnamese, and English. We also compare `WikiMatch` to state-of-the-art techniques from data integration [3] and Information Retrieval [20], as well as to a technique specifically designed to align infobox attributes [5]. The results show that `WikiMatch` consistently outperforms existing approaches in terms of F-measure, and in particular, it obtains substantially higher recall. We also present a case study where we show that, through the use of the correspondences derived by `WikiMatch`, a multilingual querying system is able to derive higher-quality answers.

## 2. PROBLEM DEFINITION

A Wikipedia article is associated with and describes an *entity* (or object). Let $A$ be an article in language $L$ associated with entity $E$. Among the different components of $A$, here, we are interested in its *title*; *infobox*, which consists of a structured record that summarizes important information about $E$; and *cross-language links*, URLs of pages in languages other than $L$ that describe $E$. An infobox $I$ contains a set of attribute-value pairs $\{\langle a_1, v_1 \rangle, \ldots, \langle a_n, v_n \rangle\}$.

Figure 1(a) shows the infobox of an English article with 14 attribute-value pairs. Since there is a one-to-one relationship between $I$ and its associated $E$, we use these terms interchangeably in the remainder of the paper. We define the set of attributes in an infobox $I$ as the *schema of $I$* ($S_I$).

The value $v$ of an attribute $a$ in an infobox $I$ may contain one or more hyperlinks to other Wikipedia entities. For example, in Figure 1(a), the value for the attribute *Directed by* contains a hyperlink to the entity *Bernardo Bertolucci*. We denote such a hyperlink by the tuple $h = (I, v, J)$, where $J$ is the infobox pointed to by $v$. We distinguish between hyperlinks that point to another entity in the same language (which define *relationships*) and hyperlinks that point to articles describing the same entity in different languages. We refer to the latter as *cross-language links*. We denote by $cl = (I_L, I_{L'})$ a link between the documents in languages $L$ and $L'$ which represent the same entity. These links can be found in most articles and are located on the pane to the left of the article.

An article is also associated with an *entity type $T$*. For example, the article in Figure 1(a) corresponds to the type "Film". There are different ways to determine the entity type for an article, including from the categories defined for the article; from the template defined for the infobox; or from the structure of the infobox. Given a set $\mathcal{I}_L$ of infoboxes in language $L$ associated with entity type $T$, we refer to the set of all distinct attributes in $\mathcal{I}_L$ as the *schema of $T$* ($S_T$). Given two infoboxes $I_L$ and $I_{L'}$ with type $T$ that are connected by a cross-language link, we refer to the union of the attributes in their schemas, $S_D = S_I \bigcup S_{I'}$, as a *dual-language infobox schema*. The problem we address can be stated as follows: Given two sets of infoboxes $\mathcal{I}_L$ and $\mathcal{I}_{L'}$ in languages $L$ and $L'$, respectively, such that both sets are associated with the entity type $T$ and the infoboxes in the sets are connected through cross-language links. To match $S_T$ and $S'_T$, the schemas of infoboxes in the two sets, we need to find correspondences (or matches) $\langle a, a' \rangle$ such that $a$ is an attribute of $S_I$, $a'$ is an attribute of $S_{I'}$, and $a$ and $a'$ have the same meaning.

## 3. THE WIKIMATCH APPROACH

`WikiMatch` works in three steps. First, it identifies mappings between entity types in different languages, *e.g.,* it determines that type "Film" in English corresponds to type "Filme" in Portuguese. It then computes, for each type, the similarity for all attribute pairs within and across languages. To do so, it leverages information available in Wikipedia, including: attribute values, link structure of articles, cross-language links, and an automatically-derived bilingual dictionary. As another source of similarity, `WikiMatch` uses Latent Semantic Indexing (LSI) [7] as a correlation measure. Because `WikiMatch` does not rely on string similarity functions for attribute names, it is effective even for languages that do not share words with similar roots.

Even though it is useful to consider multiple similarity sources, an important challenge that ensues is how to combine them. While searching for attribute correspondences, `WikiMatch` incrementally combines the different sources, and selects the *high-confidence* matches first, in an attempt to avoid error propagation to subsequent matches. As the last step, to improve recall, the derived correspondences are used to help identify additional correspondences for attributes that remain unmatched.

### 3.1 Matching Entity Types across Languages

There are different mechanisms to associate entities with types, including the assignment of categories to articles and template types to infoboxes. It is also possible to cluster the infoboxes and infer types based on their structure [26]. Regardless of the mechanism used, in Wikipedia, the entity type system is different for different languages, thus an important task is to identify the mappings between the types. `WikiMatch` adopts a simple approach that leverages the cross-language links. The intuition is that if a set of infoboxes belonging to entity type $T$ often link (through a cross-language link) to infoboxes of in a different language of type $T'$, then it is likely that types $T$ and $T'$ are equivalent.

### 3.2 Computing Cross-Language Similarities

Given two schemas $S_T$ and $S'_T$ for a type $T$, in languages $L$ and $L'$ respectively, our goal is to identify correspondences between attributes in these schemas (Section 2). To determine if a pair of attributes $< a, a' >$, where $a \in S_T$ and $a' \in S'_T$, forms a correspondence, we compute the *similarity* between $a$ and $a'$ by combining different sources of information, notably: value similarity, attribute-name correlation, and cross-language link structure.

**Cross-Language Value Similarity.** Because of the structural heterogeneity among infoboxes in different languages (see Appendix A), by combining their attributes in a unified schema for each distinct type, we gather more evidence that helps in the derivation of correspondences. We also collect for each attribute $a$ in an entity schema $S_T$, the set of values $v$ associated with $a$ in all infoboxes with type $T$. Value similarity for two attributes is then computed as the cosine similarity between their value vectors.

Since a concept can have different representations across languages, direct comparison between vectors often leads to low similarity scores. Thus, we use an automatically created translation dictionary to help improve the accuracy of the similarity score: whenever possible, the values are translated into the same language before their similarity is computed. Similar to Oh et al. [29], we exploit the cross-language links among articles in different languages to create a dictionary for their titles. The translation dictionary from a language $L$ to language $L'$ is built as follows. For each article $A$ in $L$ with a cross-language link to article $A'$ in $L'$, we add an entry to the dictionary that translates the title of article $A$ to the title of article $A'$.

Given an attribute $a$ with value vector $v_a$ in language $L$, an attribute $a'$ with value vector $v_{a'}$ in language $L'$, and a translation dictionary $D$, we construct the translated value vector of $a$ as follows: if a value of $v_a$ can be found in $D$, we replace it by its representation in $L'$. We denote the translated value vector of $a$ as $v_a^t$, and define the *value similarity* between $a$ and $a'$ as: $vsim(a, a') = cos(v_a^t, v_{a'})$, where the vector components are the raw frequencies ($tf$).

***Example 1.*** Given the vectors for *nascimento* and *born* respectively as: $v_a$={1963, Irlanda:1, 18 de Dezembro 1950:1, Estados Unidos:1} and $v_{a'}$={1963, Ireland:1, June 4 1975:1, United States: 2}, where the numbers after the colons indicate the frequency of each value. Translating $v_a$, we get $v_a^t$ ={1963, Ireland:1, December 18 1950:1, United States:1}. Thus, $vsim(v_a, v_{a'}) = cos(v_a^t, v_{a'}) = 0.71$ ■

**Link Structure Similarity.** Attribute values in an infobox often link to other articles in Wikipedia. For example, attribute *Directed by* in Figure 1(a) has the value *Bernardo*

|    | d₁ | d₂ | d₃ | d₄ | d₅ | ... | dₙ |
|---|---|---|---|---|---|---|---|
| EN born | 1 | 0 | 1 | 0 | 1 | ... | 1 |
| EN died | 0 | 1 | 1 | 1 | 1 | ... | 1 |
| EN other names | 1 | 1 | 0 | 0 | 1 | ... | 1 |
| EN spouse | 0 | 1 | 1 | 1 | 0 | ... | 0 |
| PT cônjuge | 1 | 0 | 1 | 1 | 0 | ... | 0 |
| PT falecimento | 1 | 1 | 0 | 1 | 0 | ... | 0 |
| PT morte | 0 | 0 | 1 | 0 | 1 | ... | 1 |
| PT nascimento | 1 | 0 | 1 | 0 | 1 | ... | 1 |
| PT outros nomes | 0 | 1 | 1 | 0 | 1 | ... | 1 |

(a) Co-occurrence matrix

| LSI | vsim | lsim | Attribute Pair |
|---|---|---|---|
| 0.99 | 0.45 | 0.73 | born; nascimento |
| 0.94 | 0.91 | 0.83 | falecimento; morte |
| 0.92 | 0.65 | 0.71 | died; falecimento |
| 0.73 | 0.73 | 0.26 | spouse;cônjuge |
| 0.39 | 0.60 | 0.38 | died; nascimento |
| 0.25 | 0.68 | 0.73 | died;morte |
| 0.20 | 0.47 | 0.00 | other names; outros nomes |
| 0.12 | 0.51 | 0.54 | born; morte |
| 0.00 | 0.95 | 0.58 | nascimento; falecimento |

(b) Candidate pairs sorted by LSI

**Figure 2: Some attributes for *Actor* in Pt-En**

*Bertolucci* that links to an article for this director in English. Similarly, the value of attribute *Direção* in Figure 1(*b*) links to an article for this director in Portuguese. Because of the multilingual nature of Wikipedia, the two articles for *Bernardo Bertolucci* are linked by a cross-language link. Similar to Bouma et al. [5], we leverage this feature as another source of similarity. In this example, the link structure information helps us determine that <*Directed by,Direção*> match. We define the *link structure set* of an attribute in an entity type schema $S$ as the set of outgoing links for all of its values. Given two attributes, the larger the intersection between their link structures, the more likely they are to form a correspondence. Two values are considered equal if their corresponding landing articles are linked by a cross-language link. Let $ls(a) = \{l_a^i | i = 1..n\}$ and $ls(a') = \{l_{a'}^j | j = 1..m\}$ be the link structure sets for attributes $a$ and $a'$. The *link structure similarity lsim* between these attributes is measured as: $lsim(a, a') = cos(ls(a), ls(a'))$.

For attribute values which have links, the difference between value and link similarity lies in using Wikipedia href links in two ways: their anchor texts (*vsim*) and their target URI article names (*lsim*). Since attribute values are heterogeneous (anchor texts referring to the same entity may be different, *e.g.,* "United States" and "USA") and not all values have links, both *vsim* and *lsim* are necessary.

**Attribute Correlation.** Correlation has been successfully applied in holistic strategies to identify correspondences in Web form schema matching [15, 27, 34]. There, the intuition was that synonyms should not co-occur in a given form and therefore, they should be negatively correlated. For a given language, the same intuition holds for attributes in an infobox—synonyms should not appear together. However, for identifying cross-language correspondences, the opposite is true: if we combine the attribute names for corresponding infoboxes across languages creating a dual-language infobox schema, cross-language synonyms are likely to co-occur.

While previous works applied absolute correlation measures for all attribute pairs, we use Latent Semantic Indexing (LSI) [7]. Our inspiration comes from the CLIR literature, where LSI was one of the first methods applied to match terms across languages [20]. But while LSI has traditionally been applied to terms in free text, here we use it to estimate the correlation between schema attributes.

Let $D = \{d_i | i = 1..m\}$ be the set of dual-language infoboxes associated with entity type $T$, and $A = \{a_j | j = 1..n\}$ the set of unique attributes in $D$. In the occurrence matrix $M(n \times m)$ (with $n$ rows and $m$ columns), $M(i, j) = 1$ if attribute $a_i$ appears in dual-language infobox $d_j$, and $M(i, j) = 0$ otherwise. Each row in the matrix corresponds to the occurrence pattern of the corresponding attribute over $D$. See Figure 2(a) for an example of such a matrix. We apply the truncated singular value decomposition (SVD) [20] to derive $\widetilde{M} = U_f S_f V_f^T$ by choosing the $f$ most important dimensions and scaling the attribute vectors by the top $f$ singular values in matrix $S$. SVD causes cross-language synonyms to be represented by similar vectors: if attribute names are used in similar infoboxes, they will have similar vectors in the reduced representation. This is what makes LSI suitable for cross-language matching.

To measure the correlation between attributes in different languages, we compute the cosine between their vectors. For attributes in the same language, we take the *complement* of the cosine between their vectors, and if the attributes co-occur in an infobox, we set the LSI score to 0 as they are unlikely to be synonyms. Thus, in `WikiMatch`, the LSI score for attributes $a_p$ and $a_q$ is computed as:

$$LSI(a_p, a_q) = \begin{cases} cosine(\overrightarrow{a_p}, \overrightarrow{a_q}) & \text{if } a_p \text{ in } L \wedge a_q \text{ in } L' \\ 0 & \text{if } a_p, a_q \text{ in } I_L \text{ or } I_{L'} \\ 1 - cosine(\overrightarrow{a_p}, \overrightarrow{a_q}) & \text{if } a_p \wedge a_q \text{ in } L \text{ or } L' \end{cases}$$

For attributes in the same language, a LSI score of 1 means they never co-occur in a dual-language infobox. Consequently, they are likely to be intra-language synonyms. In contrast, for attributes in different languages, a LSI score of 1 means they co-occur in every dual-language infobox. Thus, they have a good chance of being cross-language synonyms.

Note that, as illustrated in Figure 1, corresponding infoboxes are not parallel, *i.e.,* there is not a one-to-one mapping between attributes in the two languages. As a consequence, LSI is expected to yield uncertain results for cross-language synonyms. And when rare attributes are present, the same outcome will be observed for intra-language synonyms. As we discuss in Section 4, when used in isolation, LSI is not a reliable method for cross-language attribute alignment. However, if combined with the other sources of similarity, it contributes to high recall and precision.

Advantages of using LSI for finding cross-language synonyms include: (i) all attribute names are transformed into a language-independent representation, thus there is no need for translation; (ii) external resources such as dictionaries, thesauri, or automatic translators are not required; (iii) languages need not share similar words; and (iv) LSI can implicitly capture higher order term co-occurrence [18].

We have examined other alternatives for computing attribute correlations, including the measures used in [15, 27, 34]. However, since these were defined to identify synonyms within one language, they cannot be directly applied to our problem. We have also extended them to consider co-occurrence frequency in the dual infoboxes, but as we discuss in Appendix B, LSI outperforms all of them. This can be explained in part by the dimensionality reduction achieved by SVD and the consideration of the co-occurrence patterns of LSI for attribute pairs over all dual-language infoboxes.

### 3.3 Deriving Correspondences

The effectiveness of any given similarity measure varies for different attributes and entity types. For example, two attributes may have different values and yet be synonyms, or vice versa. Thus, to derive correspondences, an important challenge is how to combine the similarity measures. We propose an `AttributeAlignment` algorithm (Algorithm 1) which combines different similarity measures in such a way that they *reinforce* each other. Given as input the set of all attributes for infoboxes that belong to a given type, it groups together attributes that have the same label, and for these, combines their values—we refer to the set of such groups as $AG$. The attribute groups in $AG$ are then paired together, and for each pair, the similarity measures are computed (Section 3.2). This step creates a set of tuples that associate similarity values with each attribute pair: ($<a_p, a_q>$, $vsim, lsim, LSI$). The tuples with a LSI score greater than a threshold $T_{LSI}$ are then added to a priority queue $P$. Intuitively, a pair of matching attributes should have a high positive correlation. However, due to the heterogeneity in the data, this correlation may be weak, so $T_{LSI}$ should be set to a low value.

The tuples in $P$ are sorted in *decreasing order* of LSI score. The goal is to prioritize matches that are more likely to be correct and avoid the early selection of incorrect matches, which can result in error propagation to future matches. The similarities for a pair of attributes $a_p, a_q$ are combined as follows: If $max(vsim(a_p, a_q), lsim(a_p, a_q)) > T_{sim}$ then $<a_p, a_q>$ is a *certain candidate correspondence*. The intuition is that two attributes form a certain correspondence if they are correlated and this is corroborated by at least one of the other similarity measures. So that certain correspondences are selected early, $T_{sim}$ is set to a high value.

One potential drawback of `WikiMatch` is that it requires these two thresholds to be set. We have studied the behavior of `WikiMatch` using different thresholds, and as we discuss in Appendix B, our approach remains effective and obtains high F-measure for a broad range of threshold values.

Figure 2(a) shows a subset of the attributes in English and Portuguese for the type *Actor*. The cells in this matrix contain the number of occurrences for an attribute in each dual-language infobox. The matches in the ground truth are indicated by the arrows. Notice that *died* matches two attributes in Portuguese. Figure 2(b) shows some of the attribute pairs in $P$, with their similarity scores. For example, the pair $<born, nascimento>$ is a certain match because all similarity scores are high.

If a candidate correspondence $<a_p, a_q>$ does not satisfy the constraint in line 10 (Algorithm 1), it is added to the set of uncertain matches $U$ (line 13) to be considered later (Section 3.4). Otherwise, if it does satisfy the constraint, it is given as input to `IntegrateMatches` (Algorithm 2), which decides whether it will be integrated into an existing match, originate a new one, or be ignored. `IntegrateMatches` outputs a set of matches, $M$, where each match $m = \{a_1 \sim .. \sim a_m\}$ includes a set of synonyms, both within and across languages. `IntegrateMatches` takes advantage of the correlations among attributes to determine how to integrate the new correspondence into the set of existing matches. If neither of the attributes in the new correspondence appears in the existing matches $M$, a new matching component is created (line 5). If at least one of the attributes is already in a match $m_j$ in $M$, e.g., suppose $a_p$ appears in $m_j$, and the LSI score between $a_q$ and all attributes $a_j$ in $m_j$ is greater than the correlation threshold $T_{LSI}$ (line 8), then $a_q$ becomes a new element in $m_j$ (line 9, where $+ \sim \{a_q\}$ denotes that $a_q$ is added to the existing match $m_j$), otherwise, it is ignored. The idea is to test for positive correlations between all attributes of a match to see whether it is possible to integrate the attributes in question into the existing matches. Since $T_{LSI}$ is set low, the requirement of having positive correlations with all attributes in an existing match is not too strict and helps merge intra- and inter-language synonyms. We should note, however, that by relaxing this constraint (*e.g.,* to include only some of the attributes), it is possible to increase recall at the cost of lower precision.

`IntegrateMatches` is based on the algorithm used by Su et al. [34] to construct groups of Web form attributes. However, our experiments (Section 4.2) show that, attribute correlation alone, is not sufficient to obtain high F-measure scores. Further, since our correlation measures work for attribute pairs both within and across languages, as illustrated in the example below, `IntegrateMatches` can discover both intra and cross-language synonyms.

*Example 2.* Consider the attribute pairs in Figure 2(b) for type *Actor*, ordered by descending LSI scores, with $T_{LSI}$=0.1. Assume that the set of existing matches $M$ includes $m = \{died \sim falecimento\}$, and we have two candidate pairs, $p_1 = <died, morte>$ and $p_2 = <died, nascimento>$. Since the LSI score for *morte* and *falecimento* is greater than $T_{LSI}$, *morte* is integrated into $m$, *i.e.*, $m = \{ died \sim falecimento \sim morte\}$. In contrast, $p_2$ is not added to $m$ since the LSI score for *falecimento* and *nascimento* is zero as they are in the same language and co-occur often. ∎

---

**Algorithm 1** AttributeAlignment

1: **Input:** Set of infobox attributes for an entity type T
2: **Output:** Set of matches $M$
3: **begin**
4:   $M \leftarrow \emptyset, P \leftarrow \emptyset$
5:   **for** each pair $<a_p, a_q>$ such that $a_p, a_q \in AG$ **do**
6:     Compute $vsim, lsim, LSI$
7:     $P \leftarrow P \cup (<a_p, a_q>, vsim, lsim, LSI) \mid LSI > T_{LSI}$
8:   **while** $P \neq \emptyset$ **do**
9:     Choose pair $<a_p, a_q>$ with the highest $LSI$ score from $P$
10:    **if** $max(vsim(a_p, a_q), lsim(a_p, a_q)) > T_{sim}$ **then**
11:      $M \leftarrow$ **IntegrateMatches**($<a_p, a_q>, M$)
12:    **else**
13:      $U \leftarrow <a_p, a_q>$ /*buffering uncertain matches*/
14:    Remove $<a_p, a_q>$ from $P$
15:   $U' \leftarrow$ ReviseUncertain($U$)
16:   **for** each $u \in U'$ **do**
17:     $M \leftarrow$ **IntegrateMatches**($u, M$)
18: **end**

---

**Algorithm 2** IntegrateMatches

1: **Input:** candidate pair $<a_p, a_q>$, set of current matches $M$
2: **Output:** updated set of matches $M$
3: **begin**
4:   **if** *neither* $a_p$ *nor* $a_q \in M$ **then**
5:     $M \leftarrow M + \{a_p \sim a_q\}$
6:   **else if** *either* $a_p$ *or* $a_q$ *appears in* $M$
7:     /* suppose $a_p$ appears in $m_j$ and $a_q$ does not appear*/
8:     **for** each $a_j \in m_j$, s.t. $LSI_{qj} > T_{LSI}$ **do**
9:       $m_j \leftarrow m_j + (\sim \{a_q\})$
10: **end**

### 3.4 Revising Uncertain Matches

Since our alignment algorithm prioritizes high-confidence correspondences, it may miss correspondences that are correct but that have low confidence—the uncertain matches. Consider, for example, value similarity. While *born* and *morte (died)* are not equivalent, their similarity is high since they share many values and links—both attributes have values that correspond to dates and places. On the other hand, although *outros nomes* and *other names* are equivalent, their value similarity is low as they do not share values or links. Consequently, even though high value similarity provides useful evidence for deriving attribute correspondences, it may also prevent correct matches from being identified. The `ReviseUncertain` step uses the set $M$ of matches derived by `AttributeAlignment` (line 15) to identify additional matches, by reinforcing or negating the uncertain candidates (in set $U$). A challenge in this step is how to balance the potential gain in recall with a potential loss in precision. Our solution to this problem is to consider only the subset $U'$ of attribute pairs in $U$ whose attributes are highly correlated with the existing matches. To capture this, we introduce the notion of *inductive grouping score*. Let $<a, a'>$ be an uncertain correspondence in $U$, and let $C_a$ and $C_{a'}$ be the set of matched attributes co-occurring with $a$ and $a'$, respectively, in their mono-lingual schemas. The inductive grouping score between $a$ and $a'$ is the average grouping score of $a$ and $a'$ with each attribute in $C_a$ and $C_{a'}$:

$$\widetilde{g}(a, a') = \frac{1}{|C|} \sum_{c_a \in C_a, c'_a \in C'_a | c_a \sim c'_a} g(a, c_a) * g(a', c'_a)$$

where the grouping score $g$ is computed as follows:

$$g(a_p, a_q) = \frac{O_{pq}}{min(O_p, O_q)}$$

$O_p$ and $O_q$ are the number of occurrences of attributes $a_p$ and $a_q$, and $O_{pq}$ is the number of times they co-occur in the set of infoboxes. Note that the grouping score is computed for the schemas of the two languages separately. The inductive grouping score is high if $a_p$ and $a_q$ co-occur often with the attributes in the discovered matches.

The final step is to integrate revised matches (lines 16-18). We take advantage the certain matches in $M$ to validate the revised matches $U'$: `IntegrateMatches` is invoked again but this time it considers pairs with similarity lower than $T_{sim}$. Although we could first threshold on different values of $T_{sim}$, as we discuss in Section 4.2, revising uncertain matches as a separate step improves recall while maintaining high precision for a wide range of $T_{sim}$ values.

*Example 3*. Consider the attribute pairs in Figure 2(b), let M={*born*∼*nascimento*, *spouse*∼*cônjuge*} be the set of existing matches. The pairs <*other names*, *outros nomes*> and <*born,morte*> are uncertain candidates since their value similarities are lower than the threshold. If the attributes in these pairs co-occur often with *born* and *spouse*, the inductive grouping scores $\widetilde{g}$ of <*other names*, *outros nomes*> and <*born, morte*> are high, and thus, these candidate matches will be revised and added to $U'$. Since {*born*∼*nascimento*} has been identified as a match, *morte* cannot be integrated into this match because *morte* and *nascimento* are in the same language and co-occur in infoboxes (their LSI score is zero). In contrast, neither *outros nomes* nor *other names* appear in $M$, so this pair is added as a new match. ∎

## 4. EXPERIMENTAL EVALUATION

**Datasets.** We collected Wikipedia infoboxes related to movies from three languages: English, Portuguese, and Vietnamese. Our aim in selecting these languages was to get variety in terms of morphology and in the number of infoboxes. Portuguese and English share words with similar roots, while Vietnamese is very different from the other two languages; and there are significantly fewer infoboxes for the pair Vietnamese-English (Vn-En) than for Portuguese-English (Pt-En)—this is also reflected in the number of types covered by the Vietnamese infoboxes (see below). We selected Portuguese and Vietnamese infoboxes that belong to articles which have cross-language links to the equivalent English article. The dataset for the Pt-En language pair consists of 8,898 infoboxes, while there are 659 infoboxes for the Vn-En pair. Infoboxes that belong to the same entity type are grouped together (Section 3). There are 14 such groups for Pt-En, and 4 for Vn-En.

**Ground Truth.** We created the ground truth for all entity types in the dataset. A bilingual expert labeled as correct or incorrect all the correspondences containing attributes from two distinct languages. A pair of attributes $\langle a, a' \rangle$ is considered a correct alignment if $a$ and $a'$ have the same meaning. The ground truth set for the Pt-En pair has 315 alignments while the Vn-En pair has 160 alignments.

**Evaluation Metrics.** To account for the importance of different attributes and, consequently, of the matches involving them, we use weighted scores. Intuitively, a match between frequent attributes will have a higher weight. Let $\mathcal{C}$ be the set of cross-language matches derived by our algorithm; $\mathcal{G}$ be the cross-language matches in the ground truth; $S_T$ the set of attributes of entity type $T$ in language $L$; and $S'_T$ be the attributes in language $L'$ of the corresponding type of $T$. Given an attribute $a_i \in S_T$, we denote by $c(a_i)$ and $c_G(a_i)$ the set of attributes in $S'_T$ that correspond to $a_i$ in $\mathcal{C}$ and $\mathcal{G}$, respectively. Let $A_\mathcal{C}$ and $A_\mathcal{G}$ the set of attributes in $S_T$ that appear in $\mathcal{C}$ and $\mathcal{G}$, respectively. The *weighted scores* are computed as follows:

$$Precision = \sum_{a_i \in A_\mathcal{C}} \frac{|a_i|}{\sum_{a_k \in A_\mathcal{C}} |a_k|} Pr(c(a_i)) \quad (1)$$

$$Recall = \sum_{a_i \in A_\mathcal{G}} \frac{|a_i|}{\sum_{a_k \in A_\mathcal{G}} |a_k|} Rc(c(a_i)) \quad (2)$$

$$Pr(c(a_i)) = \sum_{a'_j \in c(a_i)} \frac{|a'_j|}{\sum_{a'_k \in c(a_i)} |a'_k|} * correct(a_i, a'_j) \quad (3)$$

$$Rc(c(a_i)) = \sum_{a'_j \in c_G(a_i)} \frac{|a'_j|}{\sum_{a'_k \in c_G(a_i)} |a'_k|} * correct(a_i, a'_j) \quad (4)$$

where $|a_i|$ represents the frequency of attribute $a_i$ in the infobox set; $correct(a_i, a'_j)$ returns 1 if the extracted correspondence $<a_i, a'_j>$ appears in $\mathcal{G}$ and 0 otherwise. Similar to [15], we compute precision and recall as the weighted averages over the precision and recall of each attribute $a_i$ (Eq. 1 and 2), and the precision and recall of attribute $a_i$ are also averaged by the contribution of each attribute $a'_j$ in $S'_T$ which corresponds to $a_i$ (Eq. 3 and 4). We compute F-measure as the harmonic mean of precision and recall. The intuition behind these measures is shown in Example 4.

*Example 4*. Consider $S_T = \{a_1, a_2\}$, $S'_T = \{a'_1, a'_2, a'_3\}$, and associated frequencies (0.6, 0.4) and (0.5, 0.3, 0.2). Suppose $G = \{\{a_1 \sim a'_1 \sim a'_2\}, \{a_2 \sim a'_3\}\}$, and the alignment algorithm derives $M = \{\{a_1 \sim a'_1\}, \{a_2 \sim a'_3\}\}$. We have $c(a_1) = \{a'_1\}$, $c(a_2) = \{a'_3\}$, while $c_G(a_1) = \{a'_1, a'_2\}$, $c_G(a_2) = \{a'_3\}$. Therefore:
$pr(c(a_1)) = \frac{0.5}{0.5} * correct(a_1, a'_1) = 1$ and $pr(c(a_2)) = 1$;
$Precision = \frac{0.6}{0.6+0.4} * pr(c1) + \frac{0.4}{0.4+0.6} * pr(c2) = 1$;
$rc(c(a_1)) = \frac{0.5}{0.5+0.3} * correct(a_1, a'_1) + \frac{0.3}{0.5+0.3} * correct(a_1, a'_2)$
$= \frac{0.5}{0.8} * 1 + \frac{0.3}{0.8} * 0 = 0.625$, and $rc(c_2) = 1$;
$Recall = \frac{0.6}{0.6+0.4} * rc(c(a_1)) + \frac{0.6}{0.6+0.4} * rc(c(a_2)) = 0.775$. ∎

**Finding Matches with `WikiMatch`.** For each entity type in the two language pairs, we ran `WikiMatch` and derived a set of matches. Table 1 shows examples of such matches. Note that we are able to find alignments where an attribute in one language is mapped to two (or more) attributes in the other language. For this experimental evaluation, we configured `WikiMatch` as follows: the threshold $T_{sim}$ used for both *vsim* and *lsim* was set to 0.6; the LSI threshold ($T_{LSI}$) was set to 0.1. The same values were used for all languages and entity types without any special tuning.

Table 1: Some alignments identified by WikiMatch

| Type | Portuguese-English | Vietnamese-English |
|---|---|---|
| Movie | direção ~ directed by<br>idioma original ~ language<br>elenco original ~ starring<br>roteiro ~ written by<br>lançamento ~ release date | đạo diễn ~ directed by<br>ngôn ngữ ~ language<br>diễn viên ~ starring<br>kịch bản ~ written by<br>kịch bản ~ story by |
| Actor | nascimento ~ born<br>data de nascimento ~ born<br>falecimento ~ died<br>morte ~ died<br>outros nomes ~ other names | nơi sinh ~ born<br>vai trò ~ occupation<br>công việc ~ occupation<br>chồng ~ spouse<br>tên khác ~ other names |

## 4.1 Comparison against Existing Approaches

We compared `WikiMatch` to techniques for schema matching, cross-language information retrieval, and to a system designed to align and complete Wikipedia Templates across languages. They are described below.

−*LSI.* We use *LSI* [7] as a technique for cross-language attribute alignment. LSI similarity scores were computed for all attribute pairs $\{a_p, a_q\}$ in an entity type $T$, where $a_p \in L$ and $a_q \in L'$. The top 1, 3, 5, and 10 scoring correspondences for each $a_p$ were used to identify matches. The best F-measure value was obtained by the top-1 configuration.

−*Bouma.* This approach for aligning infobox attributes across languages uses attribute values and cross-language links [5] (see Section 6). The input to Bouma was the same provided to `WikiMatch`, *i.e.,* attributes grouped by their entity types.

−*COMA++.* This schema matching framework supports both name- and instance-based matchers. We ran COMA++ with three configurations: name matching; instance matching; and a combination of both. To emulate approaches used in cross-language ontology alignment [10, 12], we tested a variation of COMA++ where Google Translator [14] and our automatically generated dictionary (Section 3.2) were used to translate attribute labels and values, respectively. The best configuration for Pt-En uses translation for both attribute names and values. For Vn-En, translating only the values provided the best results.[1]

**Effectiveness of `WikiMatch`.** Table 2 shows the results of the evaluation measures for the alignments derived by the different approaches applied to all entity types in our datasets. Here, we show only the results for the configurations that led to the highest F-measure (see Appendix C for the results of other configurations). In Table 2, the last row for each language pair shows the average across all types. The highest scores for each type/metric are shown in bold.

`WikiMatch` obtained the highest F-measure values for almost all types and language pairs. Its recall is lower than Bouma's for *film* in Pt-En, because it missed correct matches involving rare attributes, which occur in less than 0.5% of the infoboxes. In terms of precision, Bouma and COMA++ outperformed `WikiMatch` for some types. Still, considering the results averaged across all entity types, we tie in precision for Vn-En and come very close for Pt-En. By appropriately setting the thresholds, our approach can be tuned to obtain higher precision. However, since one of our goals is to improve recall for multilingual queries (see Section 5), where having more matches leads to the retrieval more relevant answers, we aim to obtain a balance between recall and precision.

---

[1]We also experimented with different similarity thresholds and selected the values that led to the best F-measure score.

Table 2: Weighted Precision (P), Recall (R), and F-measure (F) for the different approaches.

| | \multicolumn{12}{c}{Portuguese-English} |
|---|---|---|---|---|---|---|---|---|---|---|---|---|
| Type | WikiMatch | | | Bouma | | | COMA++ | | | LSI | | |
| | P | R | F | P | R | F | P | R | F | P | R | F |
| film | 0.97 | 0.95 | 0.96 | 0.79 | **0.99** | 0.88 | **0.99** | 0.95 | **0.97** | 0.01 | 0.20 | 0.02 |
| show | **1.00** | **0.89** | **0.94** | 0.82 | 0.68 | 0.75 | 0.98 | 0.52 | 0.68 | 0.07 | 0.05 | 0.06 |
| actor | **1.00** | **0.52** | **0.68** | **1.00** | 0.24 | 0.39 | 0.70 | 0.52 | 0.60 | 0.15 | 0.26 | 0.19 |
| artist | **1.00** | **0.72** | **0.84** | **1.00** | 0.55 | 0.71 | **1.00** | 0.34 | 0.51 | 0.75 | 0.50 | 0.60 |
| channel | 0.80 | **0.69** | **0.74** | **1.00** | 0.33 | 0.50 | 0.89 | 0.56 | 0.68 | 0.26 | 0.40 | 0.32 |
| company | 0.86 | **0.87** | **0.87** | **1.00** | 0.53 | 0.69 | 0.95 | 0.70 | 0.81 | 0.67 | 0.74 | 0.71 |
| comics ch. | 0.97 | **0.87** | **0.92** | 0.99 | 0.65 | 0.79 | 0.99 | 0.77 | 0.86 | 0.37 | 0.53 | 0.43 |
| album | **1.00** | **0.93** | **0.96** | **1.00** | 0.69 | 0.82 | **1.00** | 0.77 | 0.87 | 0.56 | 0.48 | 0.52 |
| adult actor | 0.84 | **0.59** | **0.69** | **1.00** | 0.26 | 0.41 | 0.73 | 0.43 | 0.54 | 0.22 | 0.19 | 0.20 |
| book | 0.80 | **0.75** | **0.77** | 0.75 | 0.58 | 0.66 | 0.75 | 0.66 | 0.70 | 0.15 | 0.36 | 0.21 |
| episode | 0.81 | **0.90** | **0.85** | 0.86 | 0.32 | 0.47 | **1.00** | 0.38 | 0.55 | 0.09 | 0.17 | 0.12 |
| writer | **1.00** | 0.49 | **0.65** | **1.00** | 0.22 | 0.36 | **1.00** | 0.27 | 0.43 | 0.60 | **0.49** | 0.54 |
| comics | 0.92 | **0.65** | **0.76** | **1.00** | 0.13 | 0.23 | 0.91 | 0.45 | 0.61 | 0.00 | 0.00 | 0.00 |
| fictional ch. | **1.00** | 0.69 | **0.82** | **1.00** | 0.06 | 0.11 | 0.81 | **0.81** | 0.81 | 0.36 | 0.37 | 0.36 |
| Avg | 0.93 | **0.75** | **0.82** | **0.94** | 0.45 | 0.55 | 0.91 | 0.58 | 0.69 | 0.30 | 0.34 | 0.31 |
| | \multicolumn{12}{c}{Vietnamese-English} |
| Type | WikiMatch | | | Bouma | | | COMA++ | | | LSI | | |
| | P | R | F | P | R | F | P | R | F | P | R | F |
| film | **1.00** | **0.99** | **0.99** | **1.00** | **0.99** | **0.99** | **1.00** | 0.91 | 0.95 | 0.65 | 0.62 | 0.63 |
| show | **1.00** | **0.88** | **0.93** | **1.00** | 0.36 | 0.53 | **1.00** | 0.61 | 0.76 | 0.57 | 0.49 | 0.53 |
| actor | **1.00** | **0.49** | **0.66** | **1.00** | 0.28 | 0.44 | **1.00** | 0.39 | 0.56 | 0.49 | 0.35 | 0.41 |
| artist | **1.00** | **0.65** | **0.79** | **1.00** | 0.32 | 0.48 | **1.00** | 0.25 | 0.40 | 0.72 | 0.50 | 0.59 |
| Avg | **1.00** | **0.75** | **0.84** | **1.00** | 0.49 | 0.61 | **1.00** | 0.54 | 0.67 | 0.61 | 0.49 | 0.54 |

`WikiMatch` outperforms the multilingual COMA++ configurations. This indicates that the combination of machine translation and string similarity is not effective for determining multilingual matches. This observation is also supported by the low F-measure scores for the name-based matching configuration (see Appendix C).

Overall, LSI produced the worst results. This is due to the fact that it only uses co-occurrences as a source of similarity; it does not leverage other sources of similarity which are important to distinguish between correct and incorrect correspondences. In addition, while LSI performs well given parallel input, in our scenario, its effectiveness is reduced due to the heterogeneity among infoboxes in different languages (see Appendix A).

**Effect of Cross-Language Heterogeneity.** Comparing results across languages, we see that Vn-En alignments were more accurate than the Pt-En in some cases, despite the fact that English is morphologically more similar to Portuguese. The reason for this behavior is that the dual-language infoboxes for Pt-En are more heterogeneous than the ones for Vn-En. Using our gold data, we calculated the overlap between attributes for pairs of corresponding infoboxes in languages $L$ and $L'$ (Appendix A). The result of this analysis showed that the overlap is significantly higher for Vn-En. For example, for the entity type *film* the overlap is 87% for Vn-En and only 36% for Pt-En. As a result, nearly all methods did better for this type for Vn-En. We also computed the correlations for overlap and the results for the different approaches. For all approaches, the coefficients show positive correlations among overlap and the results, indicating the results tend to be better for types that are more ho-

139

mogeneous across languages. Still, `WikiMatch` outperforms other approaches for entity types with both high (*e.g., film* in Vn-En) and low overlap (*e.g., channel*).

**Limitations.** We should note that not all correct attribute pairs co-occur in the data—some will not be found in any dual-language infobox. For example, no dual-language (Pt-En) infobox contains the attributes *prêmios* and *awards* even though they are synonyms. Like other approaches, `WikiMatch` is not able to identify such matches since all similarity measures return low scores. However, these are rare matches, which as we see from the results, do not significantly compromise recall. Another limitation of our approach is that, currently, it does not support languages that do not use alphabetical characters.

## 4.2 Contribution of Different Components

We analyzed how much each component of `WikiMatch` contributes to the results by running it multiple times, and each time removing one of the components. The results, averaged over all types, are summarized in Table 3. `WikiMatch` leads to the highest F-measure values, showing that the combination of its different components is beneficial.

**WikiMatch-ReviseUncertain.** When `ReviseUncertain` is omitted, recall drops substantially while there is little or no change to precision. This underscores the importance of this step: `ReviseUncertain` leads to F-measure gains between 14% and 20% for the two language pairs. We note that the effectiveness of `ReviseUncertain` varies across the different types: types whose correspondences have low value similarity tend to benefit more from `ReviseUncertain`.

**WikiMatch-IntegrateMatches.** This configuration generates matches without the `IntegrateMatches` step, which check the pairwise correlation constraints for the attributes in a match. As we discuss below, removing this step leads to a drop in precision for both Pt-En and Vn-En. This happens because it finds some incorrect matches that have high *lsim* or *vsim* values, which in `WikiMatch` are filtered out by `IntegrateMatches`.

**WikiMatch random.** To assess the contribution of ordering candidate pairs by their LSI scores, we compared it to a random ordering, while maintaining both value and link similarity constraints to validate match candidates. As the results show, the random ordering leads to significantly lower values for both precision and recall. This indicates the LSI ordering is effective at reducing error propagation.

**WikiMatch single step.** In `WikiMatch` single step, we omit the invocation of `IntegrateMatches` (line 17 in Algorithm 1) and consider as correspondences all candidates whose *lsim* or *vsim* values are positive. The sharp decline in F-measure provides evidence that considering certain and uncertain matches separately is crucial.

**Similarity Features.** We have also studied the contribution of different similarity sources. We report the results of three variations of `WikiMatch` where each omits the use of one feature: `WikiMatch`-vsim, `WikiMatch`-lsim, and `WikiMatch`-LSI. For `WikiMatch`-LSI, the candidate pairs were sorted in decreasing order of $max(vsim, lsim)$, and validated by the constraints on just these features. The numbers indicate that value similarity is the most important feature. Without *vsim*, F-measure drops about 29% in Portuguese and 19% in Vietnamese. Link similarity has a bigger impact Vietnamese than Portuguese. As expected, this feature is likely to be more important for language pairs with more diverse morphologies. For example, link similar-ity contributes 13% in precision for Vietnamese, while for Portuguese the contribution is 1%. Without LSI, F-measure drops 12% in Portuguese and 7% in Vietnamese.

Figure 3 shows how `WikiMatch` (*WM*) and `WikiMatch` without `ReviseUncertain` (*WM\**) behave when each of the features is removed. In all cases, the recall of *WM* is higher. This confirms the importance of `ReviseUncertain`, which is able to identify additional correct matches even when `WikiMatch` is given less evidence.

**Table 3: Contribution of different components**

| Configuration | Portuguese-English | | | Vietnamese-English | | |
|---|---|---|---|---|---|---|
| | P | R | F | P | R | F |
| WikiMatch | 0.93 | 0.75 | 0.82 | 1.00 | 0.75 | 0.84 |
| WikiMatch-ReviseUncertain | 0.94 | 0.54 | 0.66 | 1.00 | 0.59 | 0.72 |
| WikiMatch-IntegrateMatches | 0.84 | 0.70 | 0.75 | 0.95 | 0.74 | 0.82 |
| WikiMatch random | 0.74 | 0.40 | 0.50 | 0.77 | 0.56 | 0.64 |
| WikiMatch single step | 0.39 | 0.89 | 0.52 | 0.56 | 0.88 | 0.64 |
| WikiMatch-vsim | 0.90 | 0.43 | 0.58 | 1.00 | 0.51 | 0.68 |
| WikiMatch-lsim | 0.92 | 0.74 | 0.82 | 0.87 | 0.70 | 0.78 |
| WikiMatch-LSI | 0.83 | 0.64 | 0.72 | 0.89 | 0.69 | 0.78 |

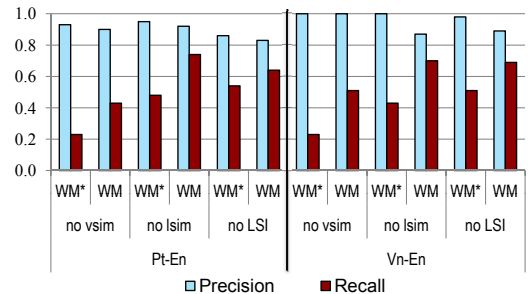

**Figure 3: Impact of ReviseUncertain**

## 5. CASE STUDY: EVALUATING CROSS-LANGUAGE QUERIES

The usual approach to answering cross-language queries is to translate the user query into the language of the articles, and then proceed with monolingual query processing. Our attribute correspondences can help retrieval systems in this translation process.

To show the benefits of identifying the multilingual attribute correspondences, below, we present a case study using WikiQuery [25], a system that supports structured queries over infoboxes. WikiQuery supports *c-queries*, which consist of a set of constraints on entity types, attribute names and values. For example, for the query: *What are the Web sites of Brazilian actors who starred in films awarded with an Oscar?*, the corresponding c-query is expressed as: *Q: Actor(born=Brazil, website=?) and Film(award=Oscar)*, where, *Actor* and *Film* are entity types; *born*, *website*, and *award* are attribute names.

The matches identified by `WikiMatch` for a given language pair are stored in a *dictionary*. To provide multilingual answers to a query, WikiQuery looks up the dictionary and retrieves, for each term in the source language, its translations into the target language. If a translation cannot be found for a given attribute $a$, the query is relaxed by removing the constraint on $a$.

**The Experiment.** We ran a set of ten c-queries (Table 4) in Portuguese and Vietnamese on the respective language datasets. We then translated the queries into English (as described above) and ran them over the English dataset. For each query, the top 20 answers were presented to two evaluators who were required to give each answer a score on a five-point relevance scale. The results were evaluated in terms of cumulative gain (CG) [16], which has been widely used in information retrieval. CG is the total relevance score of all answers returned by the system for a given query and it allows us examine the usefulness, or gain, of a result set. Figure 4 shows the CG for Portuguese queries run over the Portuguese infoboxes (*Pt*) and for Vietnamese queries run over the Vietnamese infoboxes (*Vn*); and the CG for these queries translated into English run against the English infoboxes (*Pt→En* and *Vn→En*). We can see that CG is always larger for the queries translated into English. This shows that our attribute correspondences help the translation and lead to the retrieval of more relevant answers. Because the English dataset covers a considerable portion of the contents both in Portuguese and Vietnamese infoboxes, it often returns many more answers.

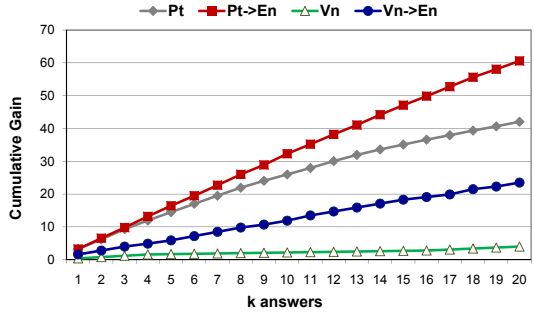

Figure 4: Cumulative Gain of $k$ answers

Table 4: List of c-queries used in the Case Study

| | |
|---|---|
| | *Movies with an actor who is also a politician* |
| 1 | filme(nome=?) and ator(ocupação="político") |
| | phim(tên=?) and diễn viên (công việc ="chính khách") |
| | *Actors who worked with director Francis Ford Coppola in a movie* |
| 2 | filme(nome=?) and ator(nome=?) and diretor(nome="francis ford coppola") |
| | phim(tên=?) and diễn viên(tên=?) and đạo diễn(tên="francis ford coppola") |
| | *Movies that won Best Picture Award and were directed by a director from England* |
| 3 | filme(direção=?) and prêmio(melhor filme=?) and diretor(nascimento\| país de nascimento\|data de nascimento="Inglaterra") |
| | phim(đạo diễn=?) and giải thưởng(phim xuất sắc nhất=?) and đạo diễn(sinh\|nơi sinh="anh") |
| | *Movies directed by director younger than 40 (born after 1970) and that have gross revenue greater than 10 million* |
| 4 | filme(receita > 10000000) and diretor(nascimento\|data de nascimento >=1970) |
| | phim(doanh thu\|thu nhập >10000000) and đạo diễn(sinh\|ngày sinh >=1970) |
| | *Books that were written by a writer born before 1975* |
| 5 | livro(nome=?) and escritor(nascimento<1975) |
| | sách(tên=?) and nhà văn(ngày sinh<1975) |
| | *Names of French Jazz artists* |
| 6 | artista(nome=?, nascimento\|país de nascimento\|país de nascimento="França", gênero="Jazz") |
| | nghệ sĩ(tên=?, sinh\|nơi sinh="Pháp", thể loại="Jazz") |
| | *Characters created by Eric Kripke* |
| 7 | personagem (nome=?, criado por="Eric Kripke") |
| | nhân vật(tên=?, sáng tác="Eric Kripke") |
| | *Names of the albums from the genre "rock" recorded before 1980* |
| 8 | album(nome=?, gênero = "Rock", gravado em <1980) |
| | album(tên=?, thể loại = "Rock", ghi âm\|thu âm <1980) |
| | *Names of artists from the genre "progressive rock" who have been born after 1950* |
| 9 | artista(nome=?, gênero = "Rock Progressivo", nascimento\|data de nascimento > 1950) |
| | nghệ sĩ(tên=?, thể loại = "Progressive Rock", sinh\|năm sinh > 1950) |
| | *Headquarters of companies with revenue greater than 10 billion* |
| 10 | companhia (sede=?, faturamento > 10 bilhões) |
| | công ty(trụ sở\|trụ sở chính=?, doanh thu\|thu nhập > 10 billion) |

Even though the CG is larger when the queries are translated into English, the gain for *Vn→En* queries is smaller than the one obtained for *Pt→En*. This is due, in part, to an artifact of our translation procedure. The Vietnamese dataset is very small, and many of the English types and attribute names do not have any correspondences in Vietnamese. As a result, the queries in our workload that include these dangling types and attribute names cannot be translated and are relaxed by WikiQuery. Although answers are returned for the relaxed queries, few (and sometimes none) of them are relevant. Since the Portuguese dataset is larger than the Vietnamese dataset, this problem is attenuated.

## 6. RELATED WORK

Cross-language matching has received a lot of attention in the information retrieval and natural language processing communities (see *e.g.,* [9, 21]). While their focus has been on documents represented in plain text, our work deals with structured information. More closely related to our work are recent approaches to ontology matching, schema matching, and infobox alignment, which we discuss below.

**Cross-Language Ontology Alignment.** Fu et al. [12] and Santos et al. [10] proposed approaches that translate the labels of a source ontology using machine translation, and then apply monolingual ontology matching algorithms. The Ontology Alignment Evaluation Initiative (OAEI) [28] had a task called *very large crosslingual resources* (VLCR). VLCR consisted of matching three large ontologies including DBpedia, WordNet, and the Dutch audiovisual archive and made use of external resources such as hypernyms relationships from WordNet and EuroWordNet—a multilingual database of WordNet for several European languages. Although related, there are important differences between these approaches and ours. While ontologies have a well-defined and clean schema, Wikipedia infoboxes are heterogeneous and loosely defined. In addition, these works consider ontologies in isolation and do not take into account values associated with the attributes. As we have discussed in Section 4, values are an important component to accurately determine matches. Last, but not least, in contrast to VLCR, our approach does not rely on external resources.

**Schema Matching.** The problem of matching multilingual schemas has been largely overlooked in the literature. The only work on this topic aimed to identify attribute correspondences between English and Chinese schemas [37], relying on the fact that the names of attributes in Chinese schemas are usually the initials of their names in PinYin (*i.e.,* romanization of Chinese characters). This solution not only required substantial human intervention and a manually constructed domain ontology, but it only works for Chinese and English. Although it is possible to combine traditional schema matching approaches [31] with automatic translation (similar to [12, 10]), as shown in Section 4, this is not effective for matching multilingual infoboxes.

Also related to our approach are techniques for uncertain schema matching and data integration. Gal et al. [4] defined a class of *monotonic* schema matchers for which higher similarity scores are an indication of more precise mappings. Based on this assumption, they suggest frameworks for combining results from the same or different matchers. However, due to the heterogeneity across infoboxes, this assumption does not hold in our scenario: matches with high similarity scores are not necessarily accurate. To this hypothesis, we have experimented with different similarity thresholds for COMA++, and for higher thresholds, we have observed a drop in *both* precision and recall.

**Cross-Language Infobox Alignment.** Adar et al. [1] proposed Ziggurat, a system that uses a self-supervised classifier to identify cross-language infobox alignments. The classifier uses 26 features, including equality between attributes and values and n-gram similarity. To train the classifier, Adar et al. applied heuristics to select 20K positive and 40K negative alignment examples. Through a 10-fold cross-validation experiment with English, German, French, and Spanish, they report having achieved 90.7% accuracy. Bouma et al. [5] designed an alignment strategy for English and Dutch which relies on matching attribute-value pairs: values $v_E$ and $v_D$ are considered matches if they are identical or if there is a cross-language link between articles corresponding $v_E$ and $v_D$. A manual evaluation of 117 alignments found only two errors. Although there has not been a direct comparison between these two approaches, Bouma et al. state that their approach would lead to a lower recall. But the superior results obtained by Ziggurat rely on the availability of a large training set, which limits its scalability and applicability: training is required for each different domain and language pair considered; and the approach is likely to be effective only for domains and languages that have a large set of representatives. Adar et al. acknowledge that because their approach heavily relies on syntactic similarity (it uses n-grams), it is limited to languages that have similar roots. In contrast, `WikiMatch` is automated—requiring no training, and it can be used to create alignments for languages that are not syntactically similar, such as for example, Vietnamese and English. Nonetheless, we would have liked to compare Ziggurat against our approach, in particular, for the Pt-En language pair. Unfortunately, we were not able to obtain the code or the datasets described in [1].

## 7. CONCLUSION

In this paper, we proposed `WikiMatch`, a new approach for aligning Wikipedia infobox schemas in different languages which requires no training and is effective for languages with different morphologies. Furthermore, it does not require external sources such as dictionaries or machine translation systems. `WikiMatch` explores different sources of similarity and combines them in a systematic manner. By prioritizing high-confidence correspondences, it is able to minimize error propagation and achieve a good balance between recall and precision. Our experimental analysis showed that `WikiMatch` outperforms state-of-the-art approaches for cross-language information retrieval, schema matching, and multilingual attribute alignment; and that it is effective for types that have high cross-language heterogeneity and few data instances. We also presented a case study that demonstrates the benefits of the correspondences discovered by our approach in answering multilingual queries over Wikipedia: by using the derived correspondences, we can translate queries posed in under-represented languages into English, and as a result, return a larger number of relevant answers.

There are a number of problems that we intend to pursue in future work. To further improve the effectiveness of `WikiMatch`, we would like to investigate the use of a fixed point-based matching strategy, such as similarity flooding [23]. Because our approach is automated, the results it produces can be uncertain or incorrect. To properly deal with this issue during the evaluation of multilingual queries, we plan to explore approaches that take uncertainty into account [8]. While in this paper we focused on infoboxes, we would like to investigate the effectiveness of `WikiMatch` on other sources of structured data present in Wikipedia.

**Acknowledgments.** We thank Gosse Bouma, Sabine Massmann and Erhard Rahm for sharing their software with us, and the reviewers for their constructive comments. Viviane Moreira was partially supported by CAPES-Brazil grant 1192/10-8. This work has been partially funded by the NSF grants IIS-0905385, IIS-0844546, IIS-1142013, CNS-0751152, and IIS-0713637.

## 8. REFERENCES

[1] E. Adar, M. Skinner, and D. S. Weld. Information arbitrage across multi-lingual wikipedia. In *WSDM*, pages 94–103, 2009.

[2] S. Auer, C. Bizer, G. Kobilarov, J. Lehmann, R. Cyganiak, and Z. G. Ives. DBpedia: A nucleus for a web of open data. In *ISWC*, pages 722–735, 2007.

[3] D. Aumueller, H. H. Do, S. Massmann, and E. Rahm. Schema and ontology matching with COMA++. In *SIGMOD*, pages 906–908, 2005.

[4] G. Avigdor. *Uncertain Schema Matching*. Morgan & Claypool Publishers, 2011.

[5] G. Bouma, S. Duarte, and Z. Islam. Cross-lingual alignment and completion of wikipedia templates. In *CLIAWS3*, pages 21–29, 2009.

[6] N. Cardoso. GikiCLEF topics and wikipedia articles: Did they blend? In *Multilingual Information Access Evaluation I. Text Retrieval Experiments*, volume 6241 of *LNCS*, pages 318–321. Springer, 2010.

[7] S. Deerwester, S. T. Dumais, G. W. Furnas, T. K. Landauer, and R. Harshman. Indexing by latent semantic analysis. *Journal of the American Society for Information Science*, 41(6):391–407, 1990.

[8] X. Dong, A. Y. Halevy, and C. Yu. Data integration with uncertainty. In *VLDB*, pages 687–698, 2007.

[9] I. Dornescu. Semantic QA for encyclopaedic questions: QUAL in GikiCLEF. In *CLEF*, pages 326–333, 2009.

[10] C. T. dos Santos, P. Quaresma, and R. Vieira. An API for multi-lingual ontology matching. In *LREC*, pages 3830–3835, 2010.

[11] S. Ferrandez, A. Toral, I. Ferrandez, A. Ferrandez, and R. Munoz. Exploiting wikipedia and eurowordnet to solve cross-lingual question answering. *Information Sciences*, 179(20):3473–3488, 2009.

[12] B. Fu, R. Brennan, and D. O'Sullivan. Cross-lingual ontology mapping - an investigation of the impact of machine translation. In *ASWC*, pages 1–15, 2009.

[13] GikiCLEF - Cross-language Geographic Information Retrieval from Wikipedia. http://www.linguateca.pt/GikiCLEF.

[14] Google translator http://www.google.com/translate.

[15] B. He and K. C.-C. Chang. Automatic complex schema matching across web query interfaces: A correlation mining approach. *ACM TODS*, 31:346–395, 2006.

[16] K. Järvelin and J. Kekäläinen. Cumulated gain-based evaluation of IR techniques. *ACM TOIS*, 20:422–446, 2002.

[17] G. Kasneci, F. Suchanek, G. Ifrim, M. Ramanath, and G. Weikum. NAGA: Searching and ranking knowledge. In *ICDE*, pages 953–962, 2008.

[18] A. Kontostathis and W. M. Pottenger. A mathematical view of latent semantic indexing: Tracing term co-occurrences. Technical report, Lehigh University, 2002.

[19] A. Kumaran, K. Saravanan, N. Datha, B. Ashok, and V. Dendi. Wikibabel: A wiki-style platform for creation of parallel data. In *ACL/AFNLP*, pages 29–32, 2009.

[20] M. Littman, S. T. Dumais, and T. K. Landauer. Automatic cross-language information retrieval using latent semantic indexing. In *Cross-Language Information Retrieval, chapter 5*, pages 51–62, 1998.

[21] B. Magnini, D. Giampiccolo, P. Forner, C. Ayache, V. Jijkoun, P. Osenova, A. Peñas, P. Rocha, B. Sacaleanu, and R. F. E. Sutcliffe. Overview of the CLEF 2006 multilingual question answering track. In *CLEF*, pages 223–256, 2006.

[22] C. D. Manning, P. Raghavan, and H. Schütze. *Introduction to Information Retrieval*. Cambridge Univ. Press, 2008.

[23] S. Melnik, H. Garcia-Molina, and E. Rahm. Similarity flooding: A versatile graph matching algorithm and its application. In *ICDE*, pages 117–128, 2002.

[24] D. Nguyen, A. Overwijk, C. Hauff, D. Trieschnigg, D. Hiemstra, and F. de Jong. Wikitranslate: Query translation for cross-lingual information retrieval using only wikipedia. In *Evaluating Systems for Multilingual and Multimodal Information Access*, volume 5706 of *LNCS*, pages 58–65, 2009.

[25] H. Nguyen, T. Nguyen, H. Nguyen, and J. Freire. Querying Wikipedia Documents and Relationships. In *WebDB*, pages 4:1–4:6, 2010.

[26] T. Nguyen, H. Nguyen, and J. Freire. Entity and relationships discovery in wikipedia. Technical report, University of Utah, Salt Lake, UT, 2010.

[27] T. Nguyen, H. Nguyen, and J. Freire. PruSM: A prudent schema matching strategy for web-form interfaces. In *CIKM*, pages 1385–1388, 2010.

[28] Ontology alignment evaluation initiative. http://oaei.ontologymatching.org.

[29] J.-H. Oh, D. Kawahara, K. Uchimoto, J. Kazama, and K. Torisawa. Enriching multilingual language resources by discovering missing cross-language links in wikipedia. In *WI-IAT- Vol.1*, pages 322–328, 2008.

[30] M. Potthast, B. Stein, and M. Anderka. A Wikipedia-based multilingual retrieval model. In *ECIR*, pages 522–530, 2008.

[31] E. Rahm and P. A. Bernstein. A survey of approaches to automatic schema matching. *VLDBJ*, 10(4):334–350, 2001.

[32] P. Schönhofen, A. Benczúr, I. Bíró, and K. Csalogány. Cross-language retrieval with wikipedia. In *Advances in Multilingual and Multimodal Information Retrieval*, pages 72–79, 2008.

[33] P. Sorg and P. Cimiano. Enriching the crosslingual link structure of wikipedia - a classification-based approach. In *WikiAI*, pages 1–6, 2008.

[34] W. Su, J. Wang, and F. Lochovsky. Holistic schema matching for web query interfaces. In *EDBT*, pages 77–94, 2006.

[35] R. Udupa and M. Khapra. Improving the multilingual user experience of wikipedia using cross-language name search. In *ACL-HLT*, pages 492–500, 2010.

[36] Vietnamese wordnet. http://vi.asianwordnet.org.

[37] H. Wang, S. Tan, S. Tang, D. Yang, and Y. Tong. Identifying indirect attribute correspondences in multilingual schemas. In *DEXA*, pages 652 –656, 2006.

[38] Wikimedia traffic analysis report. http://stats.wikimedia.org/wikimedia/squids/SquidReportPageViewsPerCountryOverview.htm.

# APPENDIX
## A. STRUCTURAL HETEROGENEITY

Using the alignments in our ground truth sets (Section 4), we analyzed the structural heterogeneity of the infoboxes by considering the overlap among attribute sets from infoboxes in a given language pair. For each infobox $I_L$ in language $L$ which has a cross-language link $cl$ to its equivalent infobox $I'_{L'}$ in language $L'$, we computed the overlap between their schemas $S_I$ and $S'_I$ as the size of intersection between attributes in $S_I$ and $S'_I$ over the size of their union. To be considered part of the intersection, an attribute pair must appear in the ground truth.

The results for each entity type and language pair are shown in Table 5. The English-Vietnamese pair is more homogeneous than the English-Portuguese pair. Considering only the entity types appearing in both language pairs (*i.e.*, *film*, *show*, *actor*, and *artist*) the average overlap is 44% for Portuguese-English and 69% for Vietnamese-English. As seen in our experimental results (Table 2), all approaches we considered have better results when the overlap is larger.

Table 5: Overlap in infoboxes

| | film | show | actor | artist | channel | company | comics ch. | album | adult actor | book | episode | writer | comics | fictional ch. |
|---|---|---|---|---|---|---|---|---|---|---|---|---|---|---|
| **Pt-En** | 36% | 45% | 42% | 52% | 15% | 31% | 59% | 52% | 47% | 38% | 31% | 63% | 47% | 32% |
| **Vn-En** | 87% | 75% | 46% | 67% | | | | | | | | | | |

## B. ADDITIONAL RESULTS

**Macro-averaging.** The weighting employed in the evaluation metrics used in Section 4 can be considered as *micro-averaging*. We also computed *macro-averaging* by discarding the weights and just counting distinct attribute-name pairs. The results in Table 6 show that `WikiMatch` is still outperforms the other approaches.

Table 6: Macro-averaging results

| | WikiMatch | | | Bouma | | | COMA++ | | | LSI | | |
|---|---|---|---|---|---|---|---|---|---|---|---|---|
| | P | R | F | P | R | F | P | R | F | P | R | F |
| PT-EN | 0.88 | **0.60** | **0.71** | **0.93** | 0.36 | 0.52 | 0.79 | 0.47 | 0.59 | 0.27 | 0.28 | 0.27 |
| VN-EN | **1.00** | 0.58 | **0.73** | **1.00** | 0.34 | 0.51 | 0.93 | 0.45 | 0.60 | 0.60 | 0.43 | 0.50 |

**Threshold Sensitivity.** We have studied the sensitivity of `WikiMatch` to variations in the thresholds used in our algorithms. Figure 5 shows the variation of the weighted F-measure as the thresholds $T_{sim}$ and $T_{LSI}$ increase. The lines show that `WikiMatch` is stable over a broad range of threshold values. As a general guideline, $T_{LSI}$ should be set low since the main purpose of LSI is to sort the candidate matches, while $T_{sim}$ should be set high as it determines the selection of the high-confidence matches. We observe a similar behavior for both language pairs: although the highest F-measure is achieved around $T_{sim} = 0.6$, the values obtained for all thresholds are comparable. The LSI score is used to sort the priority queue containing the candidate pairs. However, only attribute pairs that surpass $T_{LSI}$ are inserted into this queue. Again, the curves for $T_{LSI}$ are similar for both language pairs. F-measure changes very little for $T_{LSI}$ values between 0 and 0.6. High values of $T_{LSI}$ reduce recall and, as a consequence, F-measure also decreases.

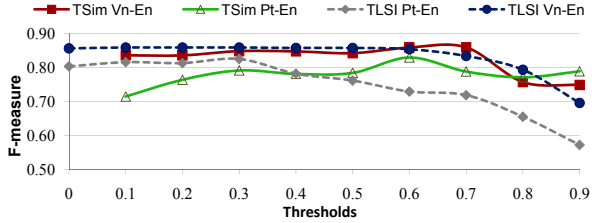

Figure 5: Impact of different thresholds

**Alternatives for Attribute Correlation.** Besides LSI, we have explored other measures for cross-language attribute correlation. Three possibilities were considered to capture the correlation between attributes $a_p$ and $a_q$: $X1 = O_{pq}$; $X2 = (1 + \frac{O_{pq}}{O_p})(1 + \frac{O_{pq}}{O_q})$; and $X3 = \frac{O_{pq} \cdot O_{pq}}{O_p + O_q}$
$O_p$ and $O_q$ are the number of occurrences of attributes $a_p$ and $a_q$ respectively, and $O_{pq}$ is the number of times they co-occur in the set of dual-language infoboxes of entity type $T$. Recall that the correlation score is used to order the candidate matches (Algorithm 1). Therefore, the best correlation measure for our approach is the one that leads to an ordering where the correct matches appear before the incorrect ones. We analyzed the ranking of matches produced by each of these measures in terms of *mean average precision* (MAP) [22], which is the standard evaluation measure for ranked items in information retrieval. It is calculated as:

$$MAP(A) = \frac{1}{|A|} \sum_{j=1}^{|A|} \frac{1}{m_j} \sum_{k=1}^{m_j} P(R_{jk})$$

where $|A|$ is the number of attributes in language $L$, $R_{jk}$ is the set of ranked pairs from the top result until attribute $a_k$, $P$ is the precision, and $m_j$ is the number of correct matches for attribute $j$. A perfect ordering ($MAP = 1$) would place all correct matches before the first incorrect match.

MAP values for LSI and the variations of the correlation score are shown in Table 7. To serve as baseline, we tried randomly ordering the attribute pairs. The results show that LSI provides the best ordering. Note, however, that all variations of $X$ are superior to random ordering. The superiority of LSI can be attributed to two factors: the dimensionality reduction brought by SVD which groups together similar infoboxes, and the fact that in addition to the co-occurrence frequency in dual-language infoboxes (which is also considered in $X1$, $X2$ and $X3$), it takes into account the occurrence pattern of the attribute pairs over the dual-language infoboxes (through the cosine distance).

Table 7: MAP for different sources of correlation

| Language Pair | LSI | X1 | X2 | X3 | Random |
|---|---|---|---|---|---|
| Portuguese-English | **0.43** | 0.26 | 0.39 | 0.35 | 0.18 |
| Vietnamese-English | **0.57** | 0.30 | 0.54 | 0.43 | 0.22 |

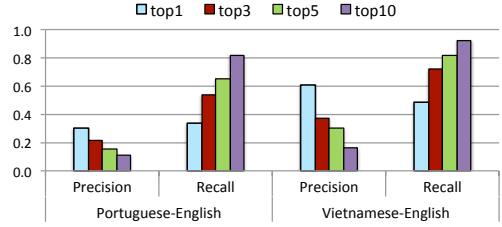

Figure 6: Top-K LSI results

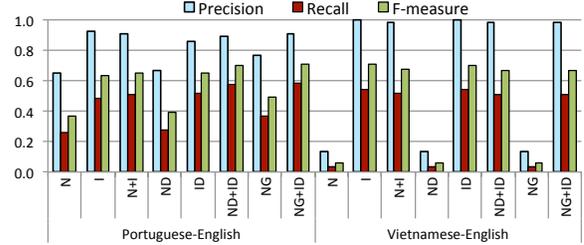

Figure 7: Different COMA++ configurations

## C. TUNING COMPETITOR SYSTEMS

In Section 4, we presented only the best configuration of each approach. Here we show the results of other configurations we tested.

**LSI.** Besides providing a source of attribute correlation in `WikiMatch`, LSI was also one of our baselines. In our experiments, we considered the *top-k* scoring matches as the alignments identified by LSI and computed the evaluation metrics. Figure 6 shows how LSI behaves for $k$ values $\in \{1, 3, 5, 10\}$. As expected, recall increases with $k$, while precision decreases.

**COMA++.** Figure 7 shows the results for different configurations of COMA++: name matcher (N), instance matcher (I), name+instance matcher (NI), using Google Translator for attribute names (N+G), and our automatically constructed dictionary for instances (I+D) and attribute names (N+D). For each configuration, we used Multiple(0,0,0) to select candidate matches as it yielded the highest F-measure. We also tried thresholds ($\lambda$) from 0 to 1 with increments of 0.1. We chose the configuration NG+ID for Pt-En, and ID for Vn-En and $\lambda = 0.01$ since these led to the highest F-measure. NG+ID had the best results in Pt-En because it combines information from more sources (names, instances, and translation). Note that the I configuration performs almost as well as the best configurations which use translation. While translation helps in some cases, in other cases an incorrect translation does more harm than good. For Vn-En, the translation of attribute names was not helpful. For instance, *dien vien* was translated to *actor* instead of *starring*, and *kinh phi* was translated to *funding* instead of *budget*. When N and I matchers are combined, the N matcher returns higher similarity scores an thus take precedence over the more reliable but lower scores of the I matcher. Therefore, NG+ID has worse results than ID only, for Vn-En. We note that even with a similarity threshold as low as 0.01, the highest recall for the best configuration of COMA++ is 0.58 for Pt-En and 0.54 for Vn-En, while for WikiMatch, at low thresholds, we obtain recall around 0.75.